\documentclass[12pt]{spieman}  % 12pt font required by SPIE;
\usepackage{amsmath,amsfonts,amssymb}
\usepackage{graphicx}
\usepackage{setspace}
\usepackage{tocloft}
\usepackage{booktabs}
\usepackage{multirow}
\usepackage{multicol}
\usepackage{makecell}
\usepackage{upgreek}
\usepackage{lineno}
\title{Radiation effects on scientific CMOS sensors for X-ray astronomy: \uppercase\expandafter{\romannumeral1}. proton irradiation}

\author[a,b]{Mingjun Liu}
\author[a,b,*]{Zhixing Ling}
\author[a,b]{Qinyu Wu}
\author[a,b]{Chen Zhang}
\author[c]{Jiaqiang Liu}
\author[b,c]{Zhenlong Zhang}
\author[a,b]{Weimin Yuan}
\author[a,b,d]{Shuang-Nan Zhang}
\affil[a]{National Astronomical Observatories, Chinese Academy of Sciences, Beijing 100101, People's Republic of China}
\affil[b]{School of Astronomy and Space Sciences, University of Chinese Academy of Sciences, Beijing 100049, People's Republic of China}
\affil[c]{National Space Science Center, Chinese Academy of Sciences, Beijing, 100190, People's Republic of China}
\affil[d]{Institute of High Energy Physics, Chinese Academy of Sciences, Beijing, 100049, People's Republic of China}

\cftpagenumbersoff{figure}
\cftpagenumbersoff{table} 
\begin{document} 
\maketitle
% Include email contact information for corresponding author
{\noindent \footnotesize\textbf{*}Zhixing Ling, \linkable{lingzhixing@nao.cas.cn} }

\begin{abstract}
Complementary metal-oxide-semiconductor (CMOS) sensors are a competitive choice for future X-ray astronomy missions. Typically, CMOS sensors on space astronomical telescopes are exposed to a high dose of irradiation. We investigate the impact of irradiation on the performance of two scientific CMOS (sCMOS) sensors between -30$^{\circ}$C to 20$^{\circ}$C at high gain mode (7.5$\times$), including the bias map, readout noise, dark current, conversion gain, and energy resolution. The two sensors are irradiated with 50 MeV protons with a total dose of 5.3$\times$10$^{10}$ p$\cdot$cm$^{-2}$. After the exposure, the bias map, readout noise and conversion gain at various temperatures are not significantly degraded, nor is the energy resolution at -30$^{\circ}$C. However, after the exposure the dark current has increased by hundreds of times, and for every 20$^{\circ}$C increase in temperature, the dark current also increases by an order of magnitude. Therefore, at room temperature, the fluctuations of the dark currents dominate the noise and lead to a serious degradation of the energy resolution. Moreover, among the 4k $\times$ 4k pixels, there are about 100 pixels whose bias at 50 ms has changed by more than 10 DN ($\sim$18 e$^-$), and about 10 pixels whose readout noise has increased by over 15 e$^-$ at -30$^{\circ}$C. Fortunately, the influence of the dark current can be reduced by decreasing the integration time, and the degraded pixels can be masked by regular analysis of the dark images. Some future X-ray missions will likely operate at -30$^{\circ}$C, under which the dark current is too small to significantly affect the X-ray performance. Our investigations show the high tolerance of the sCMOS sensors for proton radiation and prove their suitability for X-ray astronomy applications.
\end{abstract}

% Include a list of up to six keywords after the abstract
\keywords{X-ray detector; CMOS sensor; proton irradiation; sensor performance}

\begin{spacing}{2}   % use double spacing for rest of manuscript

\section{Introduction}
\label{sect:intro}  % \label{} allows reference to this section
Astronomy is an observational science that demands high sensitivity and high resolution detection of faint sources in the universe. Thanks to their high quantum efficiency and low noise\cite{2017JATIS...3c6002N}, silicon image sensors such as charge-coupled devices (CCDs), have been utilized in astronomical observations since 1980s\cite{2023arXiv230407121R}, and have become the main type of detectors for optical and X-ray astronomy. For X-ray astronomy, the ASCA\cite{1993SPIE.2006..272B} mission used CCDs as the focal plane detectors for the first time. From then on, most X-ray telescopes used CCD detectors, such as Chandra\cite{2003SPIE.4851...28G}, XMM-Newton\cite{2001A&A...365L..18S}, Swift\cite{2005SSRv..120..165B}, Suzaku\cite{2007PASJ...59S...1M}, eROSITA\cite{2021A&A...647A...1P}, and MAXI\cite{2009PASJ...61..999M}. Compared to gas detectors and multi-channel plates, CCDs have the features of high position resolution, high energy resolution, and high reliability. However, CCDs for X-ray astronomy are expensive and need to operate at low temperatures to suppress the dark current and irradiation effects\cite{2011A&A...534A..20P}, which are usually around -100$^{\circ}$C\cite{2001A&A...365L..18S,2003SPIE.4851...28G}.

In recent years, a new type of silicon detector, the CMOS sensor, has gained a wide range of applications in commercial and scientific areas, such as security systems, phone cameras and nuclear experiments, with the rapid development of semiconductor manufacturing\cite{2022JATIS...8b6001R}. Compared to CCDs, CMOS sensors have the advantages of low cost, fast readout speed, and higher working temperature \cite{2022JATIS...8b6001R}. These make CMOS detectors a competitive choice for current and future astronomical projects\cite{2009SPIE.7419E..0TH}, especially for X-ray astronomy telescopes. The Lobster Eye Imager for Astronomy (LEIA)\cite{2022ApJ...941L...2Z,2023RAA....23i5007L}, a large field-of-view X-ray instrument, used four 6 cm $\times$ 6 cm CMOS sensors as focal plane detectors. These CMOS sensors have been operating in orbit for about 1 year. The Wide-field X-ray Telescope (WXT) onboard the Einstein Probe mission (EP)\cite{2018SPIE10699E..25Y, 2022hxga.book...86Y}, which will be launched at the end of 2023, uses 48 CMOS sensors for X-ray detection.

Unlike ground-based astronomical equipment, imaging sensors in space telescopes suffer intense irradiation from all kinds of particles, typically protons and electrons\cite{2020JATIS...6a6002B}. Generally, for Si-based detectors, including CCDs and CMOS detectors, irradiation effects include ionization damage and displacement damage\cite{2001sccd.book.....J}. For ionization damage, particles ionize silicon and silicon oxide and produce electron-hole pairs. While for displacement damage, high energy particles collide with atoms in the lattice, to create defects in the bulk silicon\cite{2017ITNS...64...27B}, and then induce energy levels within the forbidden band gap and the field-enhanced emission near the defects\cite{1185167}. Extensive ground experiments and long-period performance variation of space X-ray telescopes have shown that these damage effects can lead to the degradation of the dark current, the charge transfer inefficiency (CTI), and the energy resolution of CCDs\cite{2002NIMPA.482..644A,2005SPIE.5898..201G,2020NIMPA.97864431W}. The performances of CMOS sensors, including the dark current, the readout noise, and the energy resolution, also show varying degrees of degradation under irradiation, depending on the dose and type of the irradiation\cite{7790848}. Due to the rapid development and diversity of the techniques of CMOS sensors, research on the irradiation damage on CMOS sensors is needed for current and future X-ray missions.

Protons are one of the primary particles of high-energy space radiation. For CMOS sensors, in general, an increase in dark current\cite{2020JATIS...6a6002B} and the appearance of random telegraph signals\cite{1185167} are observed after being irradiated by protons. However, the influence of proton irradiations on readout noise and energy resolution varies with the investigation conditions, including the temperature, the dose of irradiation, and the proton energy\cite{2020JATIS...6a6002B,9076701,2017ChPhB..26k4212M}. Moreover, the analysis on proton damage at the pixel level is still insufficient. Therefore, there are great demands to reveal the degradation of CMOS sensors' performances for future astronomical missions.

We started studying the CMOS sensor for X-ray detection in 2015. After a series of studies of different types of CMOS sensors\cite{2018SPIE10699E..5OW,2019JInst..14P2025W,2021JInst..16P3018L,2022JInst..17P2006H,2022JInst..17P2016W,2023PASP..135b5003W,2023NIMPA105068180W}, we proposed to use CMOS sensors for space X-ray telescopes. The CMOS sensor we use in this study is a customized sensor for X-ray astronomy, named EP4K. It has a 4k $\times$ 4k pixel array with pixel size at 15 $\upmu$m $\times$ 15 $\upmu$m and 10 $\upmu$m absorber layer. EP4K has the readout noise lower than 4 electrons with the readout frame rate at 20 Hz, described by Wu et al\cite{2022PASP..134c5006W} in detail. This sensor is designed for the WXT instrument (48 CMOSs with the total physical area of 1780 cm$^2$) onboard the EP mission for soft X-ray detection. As a new device used in space, the variation of the performance after irradiation is very important for the EP mission and other upcoming projects. Therefore, we study the performance of the CMOS sensor with the proton irradiation campaign. The experimental setup and data processing methods are described in Sec.~\ref{sect:exp}. In Sec.~\ref{sect:res&dis}, we present and discuss the degradation and uniformity of the dark current, and the influences of the proton irradiation on the noise and energy resolution. The main conclusions are summarized in Sec.~\ref{sect:dis & con}.

\section{Experimental Setup}
\label{sect:exp}

\begin{figure}
\begin{center}
\begin{tabular}{c}
\includegraphics[width=6in]{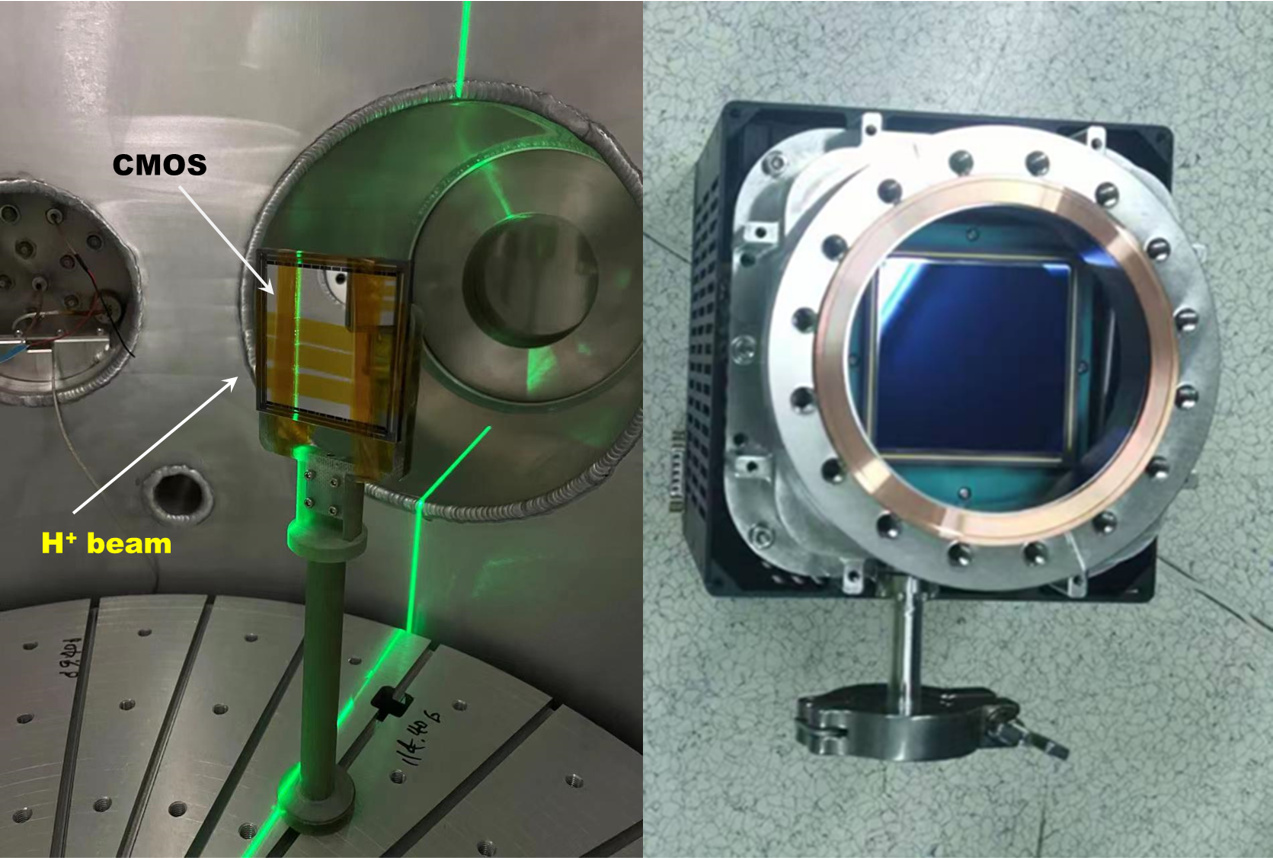}
\end{tabular}
\end{center}
\caption 
{\label{fig:equip}
The photograph of the irradiation experiment setup (left panel) and the camera used in the X-ray test (right panel).} 
\end{figure}

\begin{figure} 
\begin{center}
\begin{tabular}{c}
\includegraphics[width=6.25in]{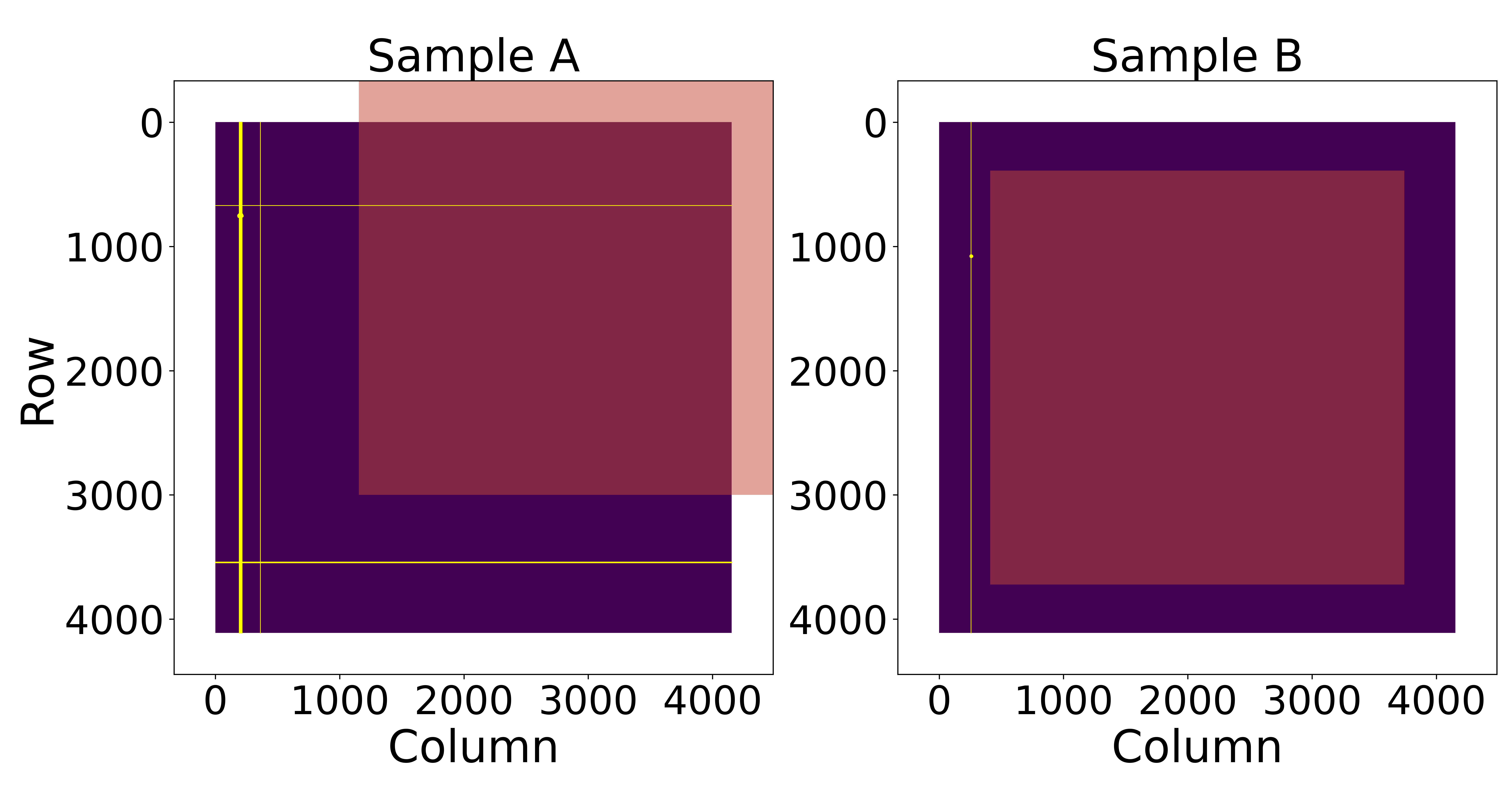}
\end{tabular}
\end{center}
\caption 
{\label{fig:radiation}
The irradiated zones (red) and defects (yellow) of Sample A (left panel) and Sample B (right panel).} 
\end{figure}

Proton irradiation experiments are carried out on the Associated Proton-Beam Experiment Platform (APEP) at the China Spallation Neutron Source (CSNS)\cite{2022NIMPA104267431L} in Dongguan in July 2022. The experimental setup is shown in the left panel of Fig.~\ref{fig:equip}. First, the H$^-$ beams are accelerated to 80 MeV in the linear accelerator (LINAC), where a few parts of the H$^-$ ions are stripped into protons, that is, associated protons, by residual gas. Then, these protons are transported to the end of LINAC and then extracted from the H$^-$ beams. Finally, the proton beams are reduced to low energy by graphite degraders and strike the target in a vacuum chamber. Two scientific EP4K CMOS sensors (Sample A and Sample B, hereafter) are placed in the focal plane without electric connection one after another. In these experiments, the energy of the proton beam is 50 MeV. The total irradiation dose is 5.3$\times$10$^{10}$ p$\cdot$cm$^{-2}$, corresponding to a 600 km altitude orbit in five years with triple margin for EP mission, calculated though OMERE\cite{2004ESASP.536..639P}. Among this, the total ionizing dose (TID) is 8.4 krad, while the total non-ionizing dose (TNID) is 2.1$\times$10$^8$ MeV$\cdot$g$^{-1}$. The irradiation dose rate is 2$\times$10$^{8}$ p$\cdot$cm$^{-2}$s$^{-1}$ and the exposure time is 270 s per sensor. The central irradiated zones are 5 cm $\times$ 5 cm with nonuniformity less than 10\% controlled by three graphite collimators with thickness of 3.5 cm\cite{2022NIMPA104267431L}. Outside the irradiated zone, the dose sharply decreases but can still introduce slight radiation effects, seen in Sections \ref{sect:res&dis}. We design different regions for the two experiments to test the variability of the performance of the pixels (the center of sensor) and the readout circuits (near the edge of sensor). As shown in Fig.~\ref{fig:radiation}, Sample A and Sample B are irradiated on the upper right zone and the central zone, respectively. 

The samples we use in the experiment are the same batch of the flight product of the WXT instrument. After selecting the flight sensors (the sensors with the fewest defects), we randomly chose two EP4K sensors in the remaining part for the proton irradiation experiment. To compare the performance of these CMOS sensors precisely, we test the sensor before and after proton irradiation. The post-irradiation characterization is carried out in September 2022. The test items include the dark current, readout noise, bias map, X-ray spectrum, and degraded pixels. The details of these results are presented in the following section.

In the tests of X-ray spectrum, the emission lines of Mn are directly released by $^{55}$Fe X-ray source. For other elements including Mg, Al, Si, Ca, and Ti, the targets are irradiated in X-ray tube to emission emit their characteristic lines.

The camera we use in this test was developed by our lab\cite{2022SPIE12191E..0LW}, as shown in the right panel of Fig.~\ref{fig:equip}. This camera could run the CMOS with its current maximum frame rate of 20 Hz and extract the X-ray events in real time. A Peltier cooler device is used to cool the detector below -30$^{\circ}$C. The camera could be connected to the X-ray beam-line using the CF100 flange. All test data are obtained at high gain mode ($7.5\times$) in a vacuum environment below $10^{-3}$ Pa.

\section{Characteristics of irradiation damages}
\label{sect:res&dis}

In our analysis, we focus on the basic performance of the CMOS sensors. Furthermore, the operation conditions in the EP satellite are especially considered, in which the CMOS sensors work at a readout frame rate of 20 Hz with an integration time of 50 ms at a temperature of -30$^{\circ}$C. There are some inherent defects for these two CMOS samples. The defects include bad rows, bad columns, and bad clusters. The defects region are shown in Fig.~\ref{fig:radiation}. Sample A has 6 bad rows, 20 bad columns, and 1 bad cluster, while Sample B has 1 bad column. Both Sample A and Sample B have a bad cluster with $\sim$10$^2$ pixels positioned on bad columns. Therefore, we exclude these defects from our data processing. 

\subsection{Bias map}
\label{sect:bias}

Bias is a basic parameter of the imaging sensor. For a CMOS sensor, the bias level is varied pixel by pixel. The bias is calculated by 50 images, and the median value for each pixel is used as the bias map. Two types of bias map corresponding to an integration time of 0 s (actually $\sim$14 $\upmu$s) and 50 ms (corresponding to 20 Hz) are compared. Although the latter should be seen as signal, its value is close to the bias due to the ultralow dark current. Therefore, `50-ms bias' is still used for convenience. The 0-s bias map reveals the performance of the on-chip electronics. The 50-ms bias map is the parameter we use for the EP-WXT.

\begin{figure} 
\begin{center}
\begin{tabular}{c}
\includegraphics[width=6.25in]{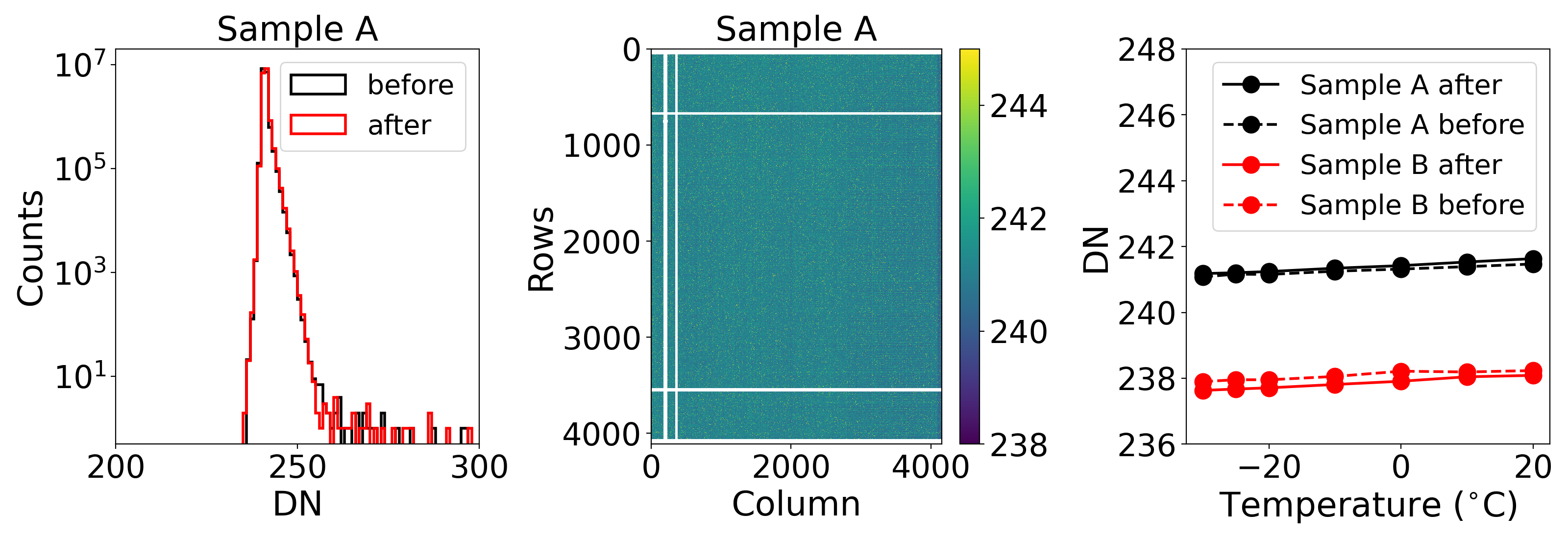}
\end{tabular}
\end{center}
\caption 
{\label{fig:bias}
The distribution of bias (left panel) at 14-$\upmu$s integration time of Sample A at -30$^{\circ}$C before (black) and after (red) proton irradiation. The exposure map after proton irradiation (middle panel) at 14-$\upmu$s integration time of Sample A at -30$^{\circ}$C. The degeneration of bias at 14-$\upmu$s integration time (right panel) of the two scientific CMOS sensors after proton irradiation. Black dashed line for Sample A before irradiation, black solid line for Sample A after irradiation, red dashed line for Sample B before irradiation, and red solid line for Sample B after irradiation.} 
\end{figure}

The bias distribution for Sample A of 0-s integration time at -30$^{\circ}$C is nearly unchanged after proton irradiation, as shown in the left panel of Fig.~\ref{fig:bias}. The black line shows the bias distribution before the proton irradiation, and the red line shows the result after the irradiation. There are no changes after irradiation. The central panel of Fig.~\ref{fig:bias} shows the image of the bias map after irradiation. No obvious structure was found inside the image. The right panel of Fig.~\ref{fig:bias} shows the median values of the bias maps before and after the irradiation test. It shows that there is no obvious change after the irradiation at each temperature both for samples A and B, represented by the black line and the red line, respectively. 

\begin{figure} 
\begin{center}
\begin{tabular}{c}
\includegraphics[width=6.25in]{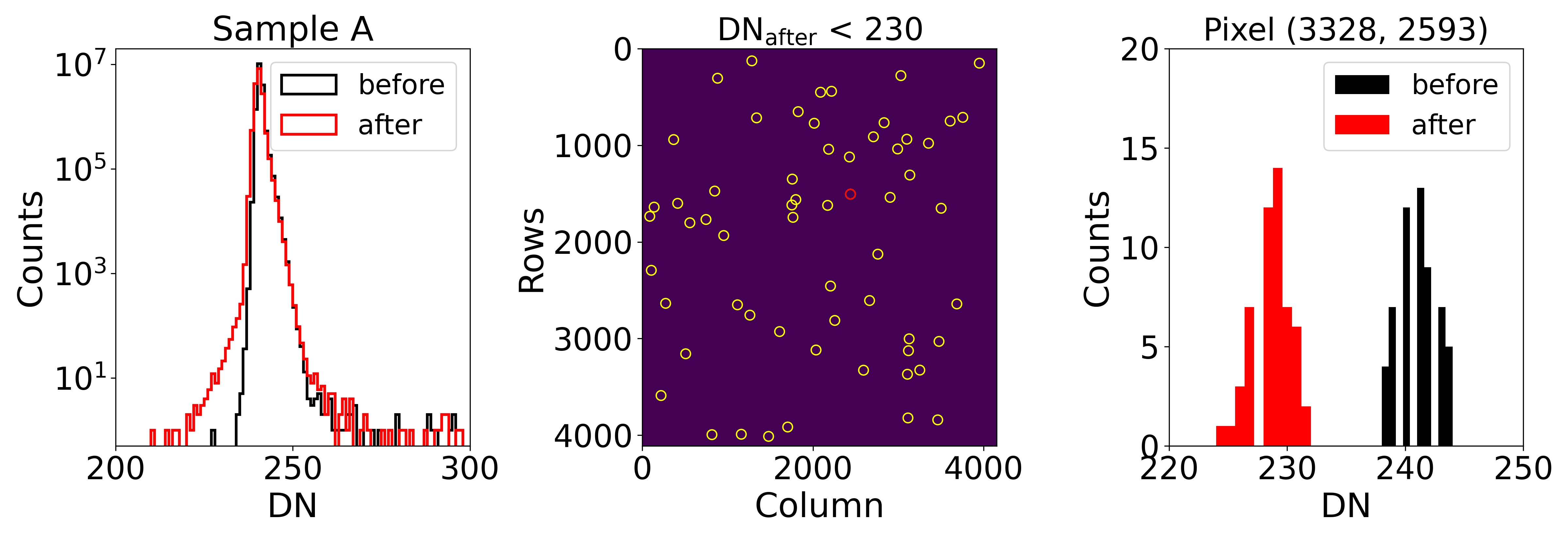}
\end{tabular}
\end{center}
\caption 
{\label{fig:bias_50}
The distribution of bias for pixels (left panel) at 50-ms integration time of Sample A at -30$^{\circ}$C before (black) and after (red) proton irradiation. The distribution of pixels (middle panel) whose DN value at 50-ms integration time and -30$^{\circ}$C less than 230 after proton irradiation for Sample A marked by yellow circle. The distribution of DN values before (black) and after (red) proton irradiation for one pixel in Sample A (right panel) whose location is marked by red circle in middle panel.}
\end{figure}

But for the bias map of 50 ms, there is a slight change. The left panel of Fig.~\ref{fig:bias_50} shows that the low side of bias distribution extends from $\sim$230 to $\sim$220 DN after proton irradiation, while the majority part of the pixels do not change. In total, about 100 pixels change after the irradiation. The locations of these pixels are shown in the middle panel of Fig.~\ref{fig:bias_50}. The distribution of these pixels on the CMOS is relatively uniform, with no clustering. The changes in bias at 50 ms for these pixels are proved by the different distributions of the readout values before and after irradiation. As shown in the right panel of Fig.~\ref{fig:bias_50}, the black and red lines show the distribution of the DN values for the bias-varied pixel (3328, 2593) before and after irradiation, respectively. Its mean value of DN obviously changed, while its readout noise only slightly increases from 3.06 e$^-$ to 3.09 e$^-$.

\subsection{Noise}

\begin{figure} 
\begin{center}
\begin{tabular}{c}
\includegraphics[width=6.25in]{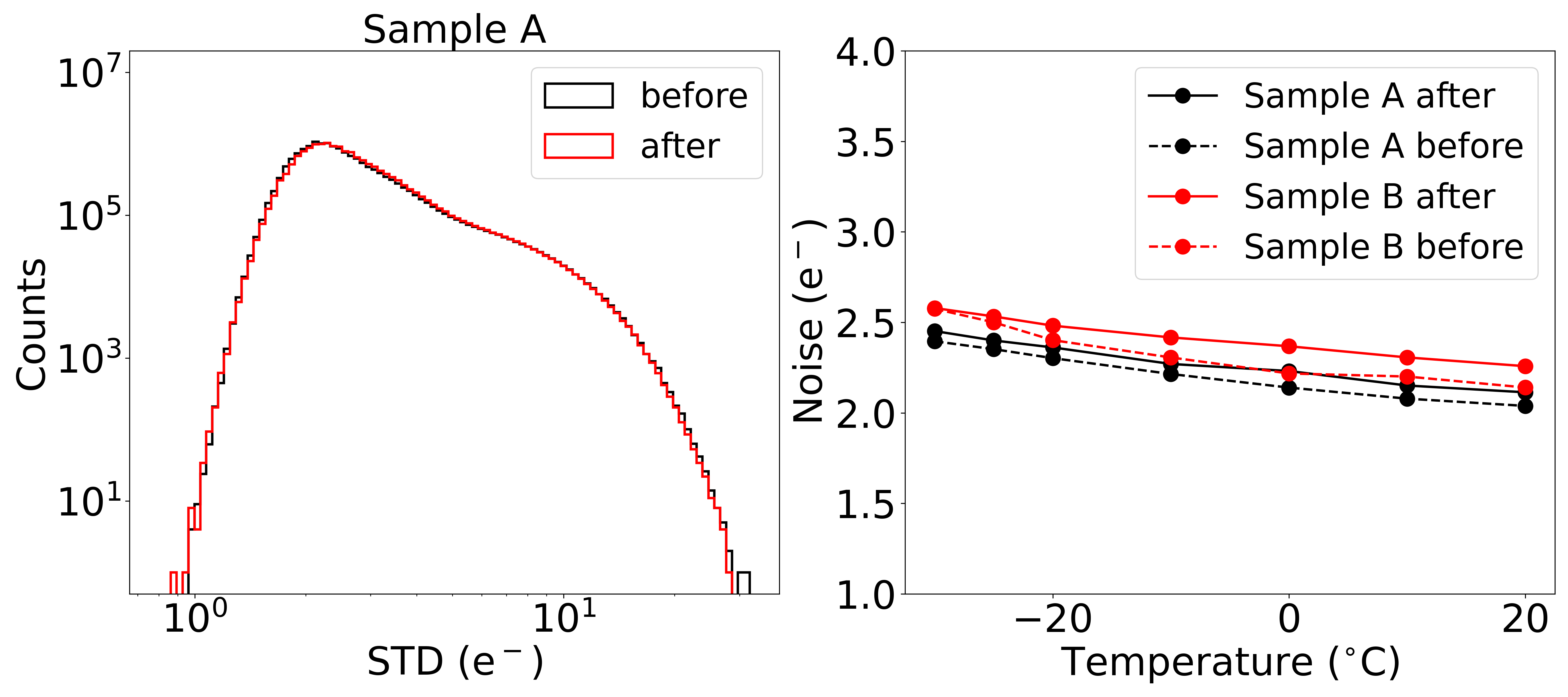}
\end{tabular}
\end{center}
\caption 
{\label{fig:noise}
The distribution of the readout noise of pixels of Sample A at 14-$\upmu$s integration time (left panel) at -30$^{\circ}$C before (black) and after (red) proton irradiation. The degradation of the readout noise (right panel) of the two scientific CMOS sensors after proton irradiation. Black dashed line for Sample A before irradiation, black solid line for Sample A after irradiation, red dashed line for Sample B before irradiation, and red solid line for Sample B after irradiation.} 
\end{figure}

The readout noise is very important for a scientific image sensor. Its variation reflects the changes in the digital part of the readout circuit. The readout noise of each pixel is calculated as the standard deviation (STD) of bias value at the minimum integration time of 14 $\upmu$s, while the readout noise of the whole CMOS is represented by the median of these STD values. The distribution of readout noise for Sample A shows no obvious change at -30$^{\circ}$C after 5.3$\times$10$^{10}$ p$\cdot$cm$^{-2}$ proton irradiation, as in the left panel of Fig.~\ref{fig:noise}; Sample B also shows the same properties. The readout noises of the whole CMOS sensor for samples A and B only increase slightly after irradiation at all temperatures from -30$^{\circ}$C to 20$^{\circ}$C, as shown in the right panel of Fig.~\ref{fig:noise}. 

\subsection{Dark current}
\begin{figure} 
\begin{center}
\begin{tabular}{c}
\includegraphics[width=6.25in]{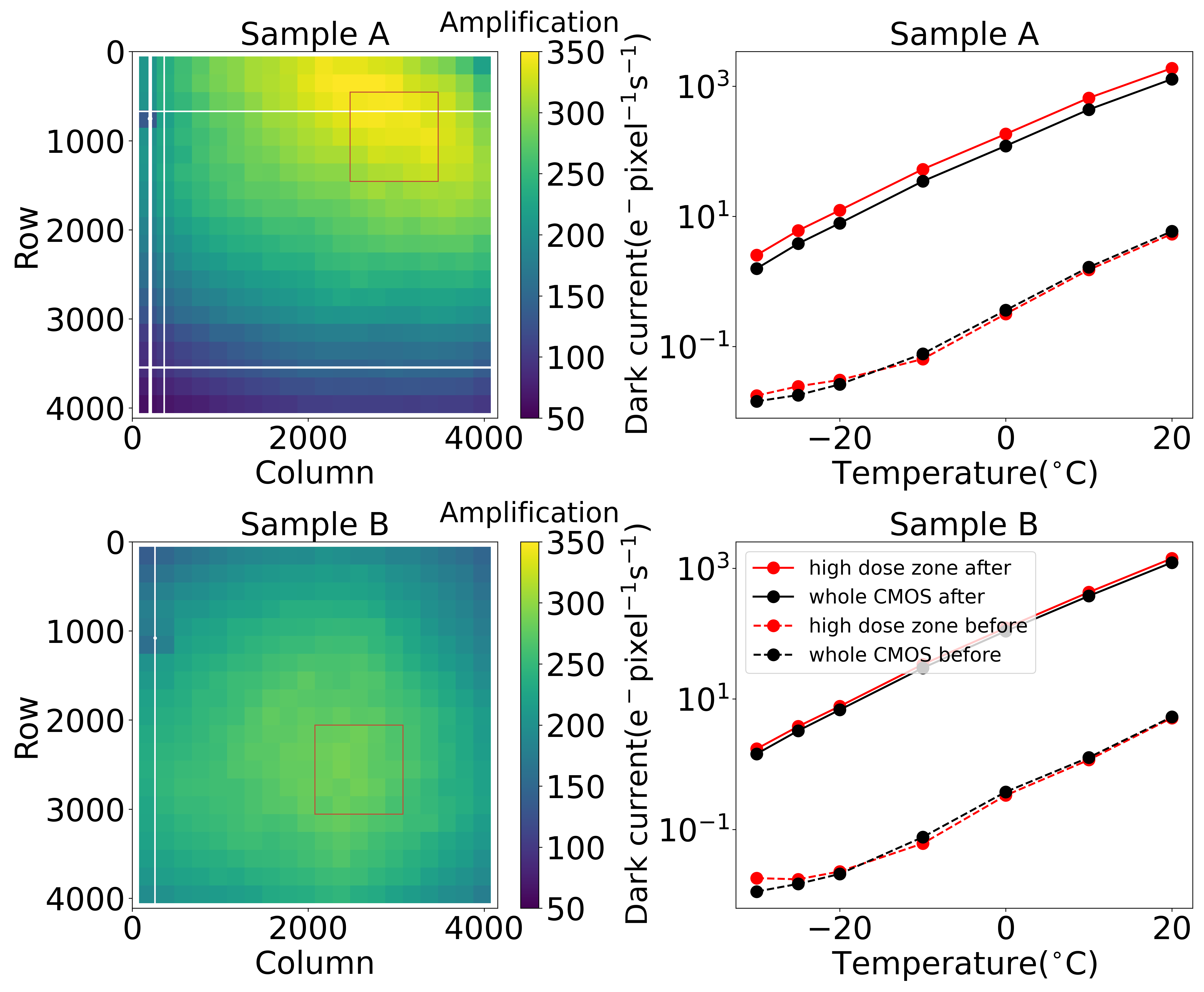}
\end{tabular}
\end{center}
\caption 
{\label{fig:dc}
The dark current ratios before and after proton irradiation at 20$^{\circ}$C of the two scientific CMOS sensors (left panel). The main irradiated zones chosen around the high degeneration regions of the two sensors are marked by red square frames.} The dark currents of the whole sensors and of the main irradiated zones at different temperatures are given in the right panel. Black dashed line for whole CMOS before irradiation, black solid line for whole CMOS after irradiation, red dashed line for main irradiated zone before irradiation, and red solid line for main irradiated zone after irradiation. 
\end{figure}

Dark current is the self-generated signal of the sensor in a dark environment, which is one of the main performance indicators for the CCD and CMOS sensors. It usually suffers from strong degradation after proton irradiation, as in many investigations mentioned in Sect.~\ref{sect:intro}. In our research, the dark currents of the whole CMOS sensor before and after irradiation are calculated using the 100-s integration time images and the longest integration time images before overexposure, respectively. The dark currents of the whole sensor at different temperatures ranging from -30$^{\circ}$C to 20$^{\circ}$C are shown in Fig.~\ref{fig:dc}. After proton irradiation, the dark current increases significantly by about 10$^2$ times. Specifically, at -30$^{\circ}$C, the dark current increases by around 110 and 128 times for samples A and B, respectively. Above -30$^{\circ}$C, the dark currents increase by 212 $\sim$ 447 times for two CMOS sensors after proton irradiation. The increase factors of the dark currents at different temperatures are similar, which shows a similar trend of increase of the dark current by about one order of magnitude for every 20$^{\circ}$C increment.

The degradation of the dark current is related to the irradiation dose. As mentioned in Sect.~\ref{sect:exp}, only part of the sensor is mainly irradiated in our research for the two CMOS sensors. Therefore, we obtain the dark current distribution on a CMOS by dividing the whole sensor into 20 $\times$ 20 zones except for the edges and computing the median in each zone. The dark current ratios before and after proton irradiation at 20$^{\circ}$C are shown as Fig.~\ref{fig:dc}. The maximum values of the dark current ratios for these 400 zones at 20$^{\circ}$C are 354 and 288 for samples A and B, respectively. The main irradiated zones of the two sensors have a greater increase of dark current after irradiation at 20$^{\circ}$C. At different temperatures, the dark current level of the main irradiated regions is always higher than the level of the whole sensor. These indicate the positive correlation between the dark current increment and the proton dose. This nonuniformity of the dark current increment relates to the nonuniformity of the incident proton beams. Therefore, the distribution of the dark current ratio in Fig.~\ref{fig:dc} means that the effective beam spot of a proton extends an ideal 5 cm $\times$ 5 cm zone as mentioned in Sec.~\ref{sect:exp}.

\subsection{Conversion gain}

\begin{figure} 
\begin{center}
\begin{tabular}{c}
\includegraphics[width=6.25in]{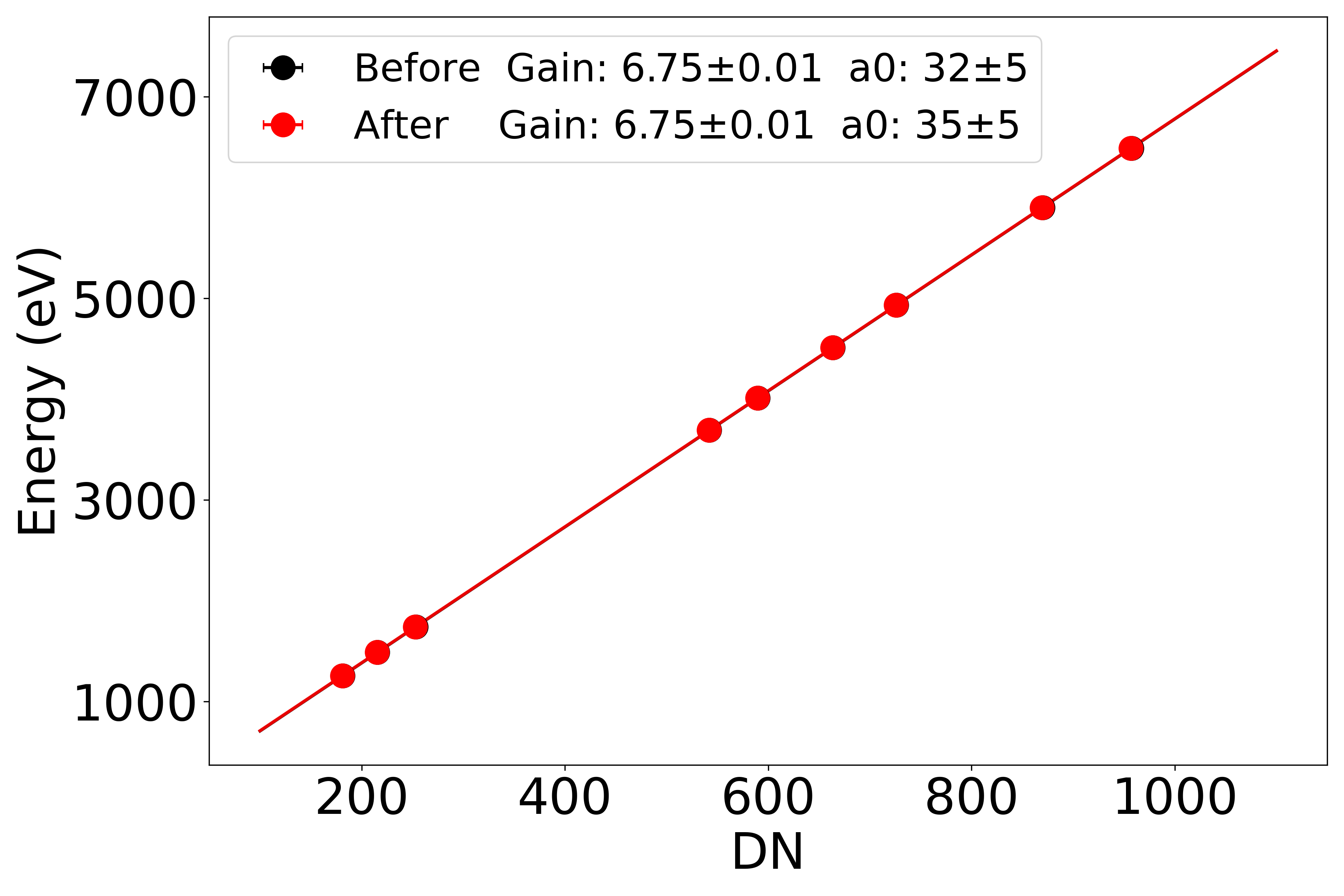}
\end{tabular}
\end{center}
\caption 
{\label{fig:gain}
The linear fit of the locations of emission lines under 50-ms integration time of Sample A before (black) and after (red) proton irradiation at -30$^{\circ}$C.} 
\end{figure}

The conversion gain (hereafter gain) quantitatively describes the conversion factor from the energy of an incident photon to the value of the digital signal recorded. The gain is obtained from the fitting of the positions of the centroid of emission lines in the X-ray spectrum with a linear function $y=a_1x+a_0$, where $x$ is the location of emission lines in DN, $y$ is the energy of the emission lines and the slope $a_1$ is the gain.  We use the emission lines of Mg, Al, Si, Ca, Ti, and Mn elements in this work. The integration time of X-ray spectrum test is 50-ms. The gains and intercepts of the two sensors at -30$^{\circ}$C and 20$^{\circ}$C are summarized in Tab.~\ref{tab:gain}. The linear fitting of Sample A at -30$^{\circ}$C is shown in Fig.~\ref{fig:gain}. The gains of the two sensors are not changed by proton irradiation at both -30$^{\circ}$C and 20$^{\circ}$C.

\begin{table}[ht]
    \caption{The conversion gains and intercepts of the two sCMOS sensors before and after proton irradiation.}
    \label{tab:gain}
    \begin{center}
    \begin{tabular}{cccccc}
    \hline\hline
    \rule[-1ex]{0pt}{3.5ex} \multirow{2}{*}[-1ex]{CMOS} & \multirow{2}{*}[-1ex]{Temperature} & \multicolumn{2}{c}{Gain (eV/DN)} & \multicolumn{2}{c}{Intercept (eV)}\\ \cmidrule(r){3-6}
    \rule[-1ex]{0pt}{3.5ex} & & before & after & before & after\\
    \hline\hline
    \rule[-1ex]{0pt}{3.5ex} \multirow{2}{*}[-1ex]{Sample A} & -30$^{\circ}$C & 6.75$\pm$0.01 & 6.75$\pm$0.01 & 32$\pm$5 & 35$\pm$5\\ \cmidrule(r){2-6}
    \rule[-1ex]{0pt}{3.5ex}  & 20$^{\circ}$C & 6.62$\pm$0.01 & 6.64$\pm$0.02 & 20$\pm$6 & -5$\pm$11\\
    \hline
    \rule[-1ex]{0pt}{3.5ex} \multirow{2}{*}[-1ex]{Sample B} & -30$^{\circ}$C & 6.64$\pm$0.01 & 6.62$\pm$0.01 & 34$\pm$5 & 43$\pm$5\\  \cmidrule(r){2-6}
     \rule[-1ex]{0pt}{3.5ex} & 20$^{\circ}$C & 6.52$\pm$0.01 & 6.51$\pm$0.02 & 24$\pm$6 & -3$\pm$11\\
    \hline\hline
    \end{tabular}
    \end{center}
\end{table}

\subsection{Energy resolution}

\begin{figure} 
\begin{center}
\begin{tabular}{c}
\includegraphics[width=6.25in]{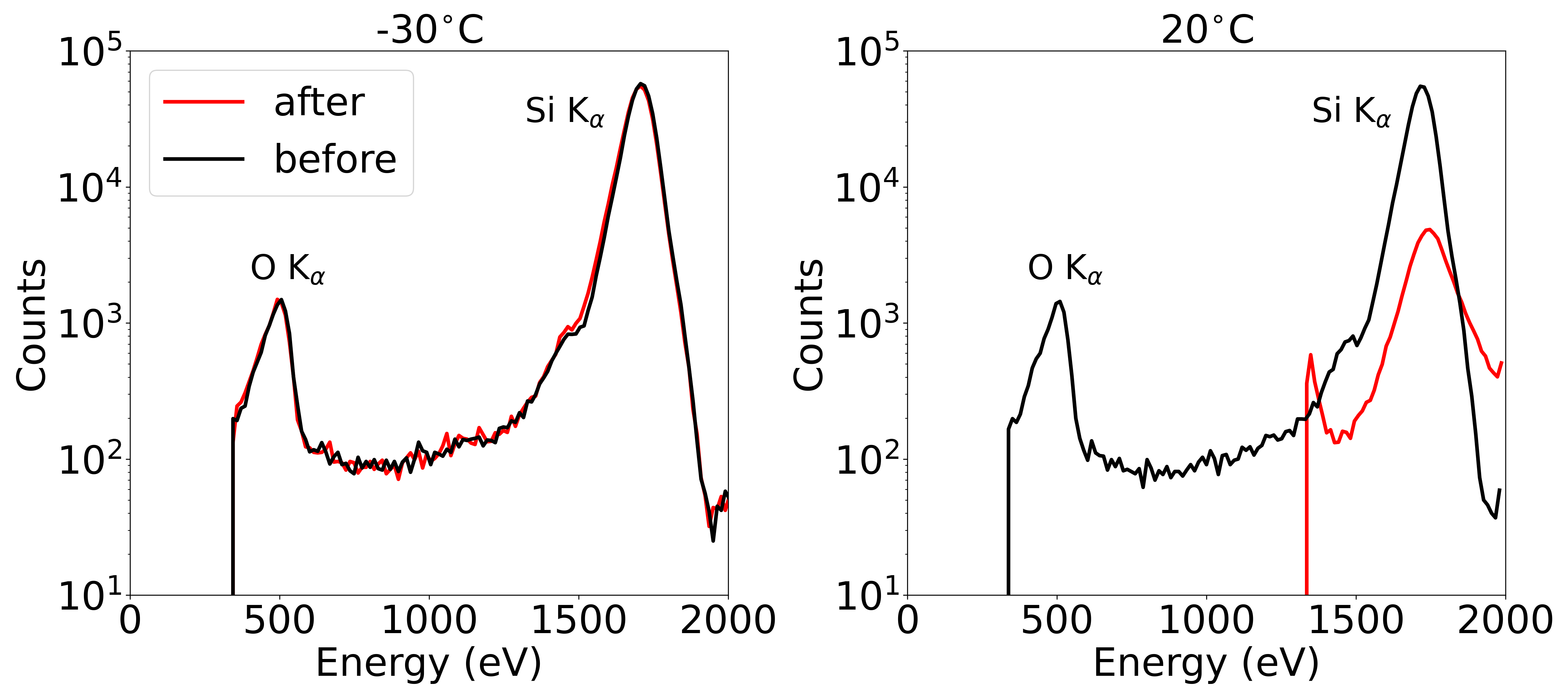}
\end{tabular}
\end{center}
\caption 
{\label{fig:FWHM}
The spectrum of the emission lines for Si under 50-ms integration time of Sample A at -30$^{\circ}$C (left panel) and 20$^{\circ}$C (right panel) before (black line) and after proton irradiation (red line). The left peak is the O K$_{\alpha}$ line. The lowest threshold of energy at 20$^{\circ}$C after proton irradiation is set as 200 DN due to the large dark current.} 
\end{figure}

The energy resolution describes the spectrum quality and is important for astronomical applications. The full widths at half maximum (FWHM) of the emission lines represent the energy resolutions of the detector. The FWHM is obtained by the Gaussian fit of the emission lines in the spectrum from single pixel events with 50-ms integration time. The FWHMs before and after irradiation are summarized in Tab.~\ref{tab:FWHM}. The spectra of Si K$_{\alpha}$ line of Sample A are given in Fig.~\ref{fig:FWHM}. At a low temperature, the energy resolution does not show obvious change after irradiation for the two sCMOS sensors, showing a good performance that can satisfy the requirement of the EP mission. But at a higher temperature, the degradation of the energy resolution is severe. For example, the FWHM of Si K$_{\alpha}$ of Sample A changes from 93.9 eV to 141.6 eV, as shown in Tab.~\ref{tab:FWHM} and Fig.~\ref{fig:FWHM}. Since the dark current at 20$^{\circ}$C after proton irradiation is around 1200 e$^-$pixel$^{-1}$s$^{-1}$ corresponding to $\sim$60 e$^-$pixel$^{-1}$ for each image, the energy resolution suffers serious degradation. These have already dominated the readout noise as shown in Fig.~\ref{fig:noise}, leading to the degradation of the energy resolution.

\begin{table}[ht]
    \caption{The energy resolutions (FWHM, in unit of eV) of the two sCMOS sensors before and after proton irradiation.}
    \label{tab:FWHM}
    \begin{center}
    \begin{tabular}{lcccc}
    \hline\hline
    \rule[-1ex]{0pt}{3.5ex} \multirow{2}{*}[-1ex]{-30$^{\circ}$C} & \multicolumn{2}{c}{Sample A} & \multicolumn{2}{c}{Sample B}\\ \cmidrule(r){2-5}
    \rule[-1ex]{0pt}{3.5ex} & before & after & before & after\\
    \hline
    \rule[-1ex]{0pt}{3.5ex}  Si K$_{\alpha}$ & 98.2$\pm$2.3 & 102.1$\pm$2.5 & 101.8$\pm$2.4 & 106.5$\pm$2.9\\
    \hline
    \rule[-1ex]{0pt}{3.5ex}  Ti K$_{\alpha}$ & 148.3$\pm$1.2 & 153.1$\pm$1.6 & 152.7$\pm$1.4 & 156.4$\pm$1.8\\
    \hline
    \rule[-1ex]{0pt}{3.5ex}  Ti K$_{\beta}$ & 160.8$\pm$1.8 & 164.8$\pm$2.0 & 166.4$\pm$2.2 & 169.5$\pm$2.4\\
    \hline\hline
    \rule[-1ex]{0pt}{3.5ex} \multirow{2}{*}[-1ex]{20$^{\circ}$C} & \multicolumn{2}{c}{Sample A} & \multicolumn{2}{c}{Sample B}\\ \cmidrule(r){2-5}
    \rule[-1ex]{0pt}{3.5ex} & before & after & before & after \\
    \hline
    \rule[-1ex]{0pt}{3.5ex}  Si K$_{\alpha}$ & 93.9$\pm$2.3 & 141.6$\pm$3.1 & 98.1$\pm$2.8 &  142.8$\pm$4.6\\
    \hline
    \rule[-1ex]{0pt}{3.5ex}  Ti K$_{\alpha}$ & 144.4$\pm$1.4 & 197.1$\pm$3.4 & 147.9$\pm$1.6 & 198.8$\pm$4.1\\
    \hline
    \rule[-1ex]{0pt}{3.5ex}  Ti K$_{\beta}$ & 156.4$\pm$1.9 & 247.0$\pm$6.8 & 160.8$\pm$2.2 &  235.2$\pm$4.8\\
    \hline\hline
    \end{tabular}
    \end{center}
\end{table}

\subsection{Degraded pixel}

\begin{figure}
\begin{center}
\begin{tabular}{c}
\includegraphics[width=6.25in]{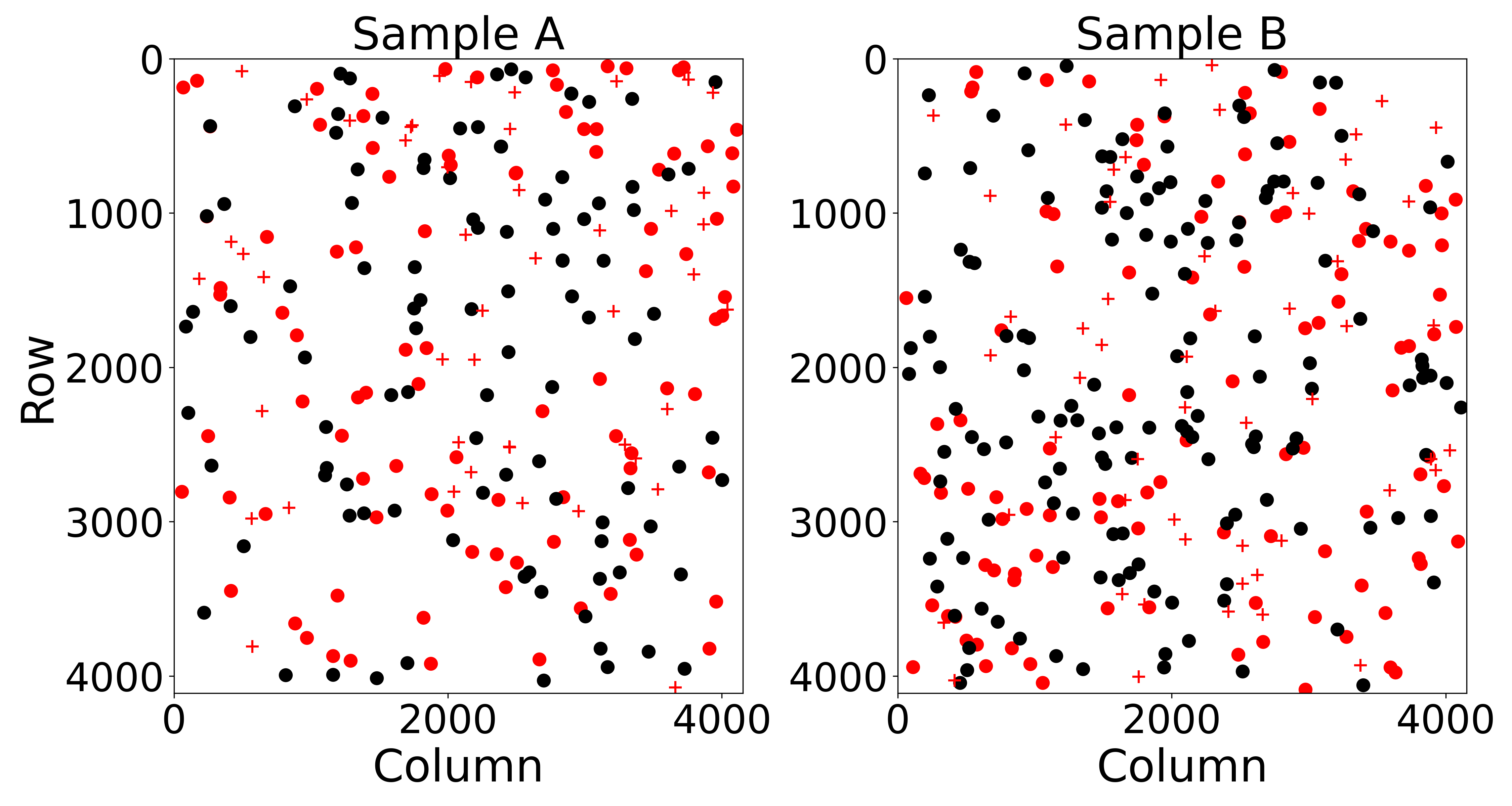}
\end{tabular}
\end{center}
\caption 
{\label{fig:pixel}
The degraded pixels at -30$^{\circ}$C after irradiation for the two sCMOS sensors. Red point for bias increase by over 10 DN, black point for bias decrease by over 10 DN, red '+' for noise increase by over 15 e$^-$.} 
\end{figure}

At -30$^{\circ}$C, although the performance of the whole sensor does not change much after irradiation, there are still a few newly generated defect pixels (degraded pixels hereafter) whose bias, dark current, and readout noise are significantly degraded. These degraded pixels are defined as pixels whose bias changes by over 10 DN under 50-ms integration time or whose noise increases by over 15 e$^-$ at -30$^{\circ}$C after proton irradiation. The positions of these degraded pixels are shown in Fig.~\ref{fig:pixel}. After proton irradiation, 94 pixels in Sample A and 108 pixels in Sample B have a DN increment by over 10, while 104 pixels in Sample A and 146 pixels in Sample B have a DN decrease by over 10. For the readout noise, the numbers of pixels whose noise increases by over 15 e$^-$ are 48 and 54 for samples A and B, respectively. The distributions of these degraded pixels are uniform and do not show clusters in excess of 9 pixels. In addition, pixels with large varying bias are hardly coincident with pixels with large varying noise, indicating the low correlation between the changes in bias and the changes in noise.

\begin{figure}
\begin{center}
\begin{tabular}{c}
\includegraphics[width=6.25in]{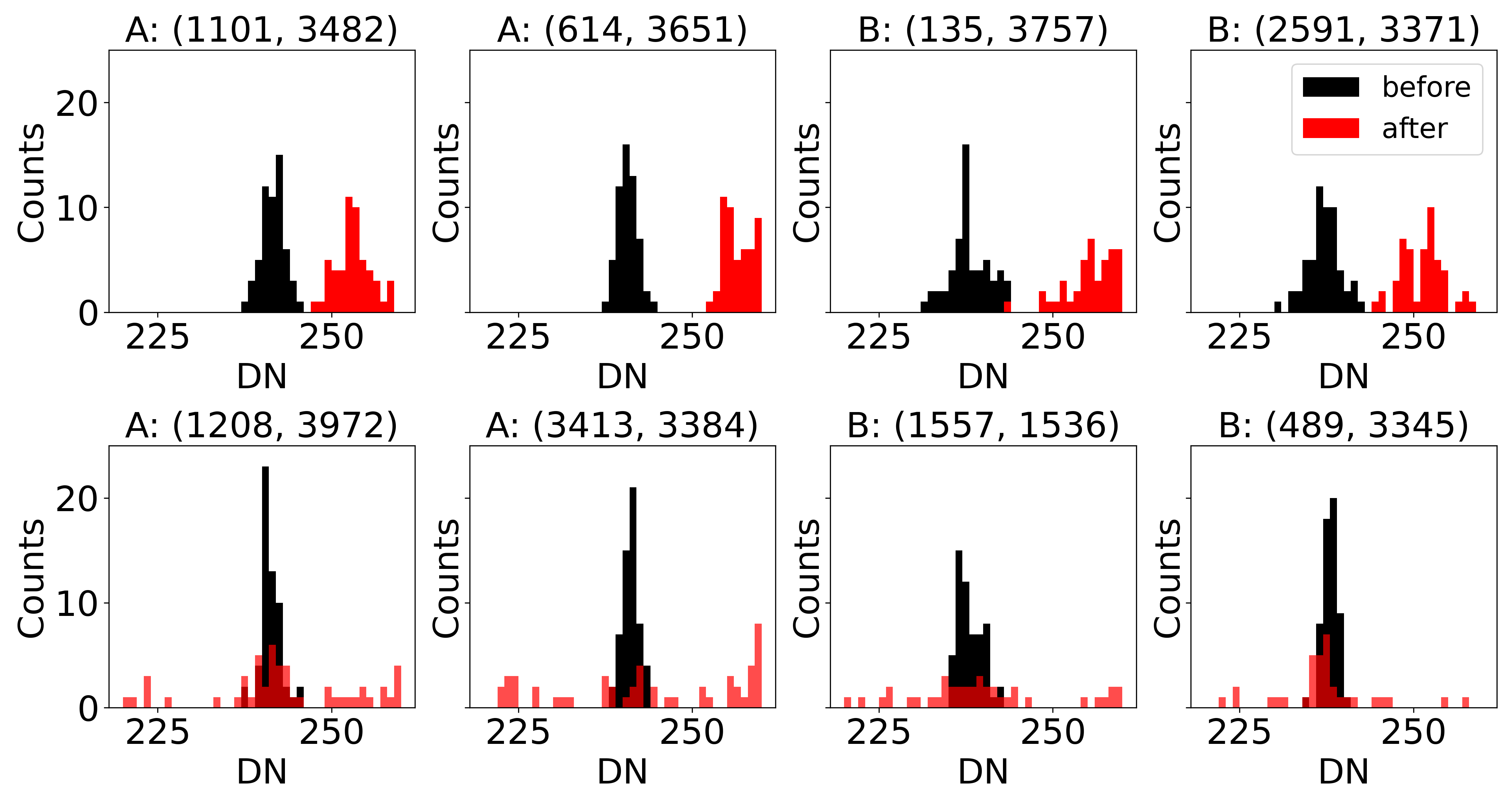}
\end{tabular}
\end{center}
\caption 
{\label{fig:pixel_hist}
The distribution of the bias values of several degraded pixels on the two sCMOS sensors under an integration time of 50 ms before (black) and after (red) proton irradiation. Upper panel: pixels with bias values increment by over 10 DN. Bottom panel: pixels with noise increment by over 15 e$^-$.} 
\end{figure}

To study those degraded pixels carefully, we extract the DN values in each frame of these pixels before and after irradiation to understand their properties. The distributions of the bias values under an integration time of 50 ms for some of these pixels are shown in Fig.~\ref{fig:pixel_hist}. For pixels whose median bias value increases by over 10 DN (upper panel of Fig.~\ref{fig:pixel_hist}), the histogram gives similar distributions centered on lower or higher DN values before and after irradiation, respectively. This is also true for pixels with a decreased bias value, which have been discussed in Sec.~\ref{sect:bias} and has been shown in Fig.~\ref{fig:bias_50}. For pixels whose noise increases by over 15 e$^-$ (bottom panel of Fig.~\ref{fig:pixel_hist}), their DN distributions usually have multipeaks without significant extra bias after proton irradiation, which may originate from the random telegraph signal. There is no significant correlation between the changes in bias values and the changes in noises.

\section{Discussion and conclusion}
\label{sect:dis & con}

Scientific CMOS sensors have great potential for future astronomical projects, especially X-ray missions. The intense irradiation of high-energy particles significantly influences the performance of the sensors of space-based telescopes. However, irradiation damage to sCMOS sensors for space missions is a subject of ongoing research. In this work, we studied many sensor characteristics, including the dark current, the readout noise, the bias map, the X-ray spectrum and the degraded pixels of two scientific CMOS sensors after the irradiation of protons of 50 MeV with a total dose of 5.3$\times$10$^{10}$ p$\cdot$cm$^{-2}$ from -30$^{\circ}$C to -20$^{\circ}$C. For CCD sensors, the displacement effect of proton irradiation increases the dark current and trap charge, leading to the degradation of the CTI. However, without the charge transfer process, CMOS sensors show better resistance to irradiation damage. From our experiments, the performances, except for the dark current, of the two sCMOS sensors hardly change after proton irradiation for an operating temperature of -30$^{\circ}$C.

The bias distributions under an integration time of 14 $\upmu$s remain unchanged for all temperatures. The bias distribution under an integration time of 50 ms slightly extends its low side from $\sim$230 to $\sim$220 DN after irradiation, containing around 100 pixels. The biases of these pixels decrease, while their noises remain unchanged.

The readout noises of the two sensors are not significantly degraded at all temperatures with a slight increase by around $\sim$0.1 e$^-$ and a similar distribution after irradiation.

The dark currents of the sensors increase by 110$\sim$447 times after irradiation, varying with temperature. While the main irradiation zone shows a more serious degradation on the dark current, which shows a positive correlation with the dose of the irradiation. The dark current increases by one order of magnitude for every 20$^{\circ}$C increment. At 10$^{\circ}$C, the dark current is around 400 e$^-$pixel$^{-1}$s$^{-1}$ after irradiation, corresponding to $\sim$2 e$^-$ per frame for the readout speed of 20 Hz, which is comparable to the readout noise (2.0$\sim$2.7 e$^-$). Thus, for the applications of sCMOS sensors at room temperature, the integration time should be designed to be shorter to mitigate the impact of the dark current. For the EP mission, with a working temperature of -30$^{\circ}$C and an integration time of 50 ms, the degraded dark currents can hardly impact the performance of the sensor.

The conversion gains remain unchanged at both -30$^{\circ}$C and 20$^{\circ}$C. Energy resolutions are also unchanged at -30$^{\circ}$C. These indicate that this CMOS sensor can perform well under intense irradiation and is suitable for applications on space missions like the EP satellite. However, at 20$^{\circ}$C, the energy resolutions suffer serious degradation due to the large dark current mentioned above.

Finally, a few pixels exhibit an anomalously large amount of degradation after proton irradiation. With the pixel-by-pixel survey for these degraded pixels, two relatively independent degradation patterns by proton radiation are found, the bias shift and readout noise increment. Fortunately, these pixels are very few. For 4k $\times$ 4k pixels on the whole CMOS sensor, only $\sim$100 pixels have a large change of bias by more than 10 DN under an integration time of 50 ms, and only $\sim$10 pixels have a large change of noise by more than 15 e$^-$ at -30$^{\circ}$C after proton irradiation. These two types of degraded pixels do not overlap, indicating that there is no significant correlation between changes in bias values and changes in noises. For applications on the EP satellite, these degraded pixels can be masked out by regular data analysis of dark exposures.

These performances show that this large-format sCMOS sensor, EP4K, has a high tolerance for proton radiation and is suitable for future X-ray missions.

% \disclosures 
%\section*{Disclosures}
%All authors of this work have no relevant financial interests in the manuscript and no other potential conflicts of interest.

%Conflicts of interest should be declared under a separate header, above Acknowledgments. If the authors have no competing interests to declare, then a statement should be included declaring no conflicts of interest. For assistance generating a disclosure statement, see the form available from  the International Committee of Medical Journal Editors website: \linkable{http://www.icmje.org/conflicts-of-interest/} 

\section* {Code, Data, and Materials Availability} 
The code and data underlying this article will be shared on reasonable request
to the corresponding author.

\section*{Acknowledgments}
This work is supported by the National Natural Science Foundation of China (grant No. 12173055) and the Chinese Academy of Sciences (grant Nos. XDA15310000, XDA15310100, XDA15310300, XDA15052100).

%%%%% References %%%%%

\bibliographystyle{spiejour}   % makes bibtex use spiejour.bst
\bibliography{ref}   % bibliography data in report.bib

%The references are numbered in the order of their first citation. Citations to the references are made using superscripts, as demonstrated in the preceding paragraph. One may also directly refer to a reference within the text, for example, ``as shown in Ref.~\citenum{Metropolis53} ...''  Two or more references should be separated by a comma with no space between them. Multiple sequential references should be displayed with a dash between the first and last numbers \cite{Alred03,Perelman97,Lamport94,Goossens97,Metropolis53}. 

%%%%% Biographies of authors %%%%%

%\vspace{1ex}
%\noindent Biographies and photographs of the authors are not available.

%A brief professional biography of approximately 75 words may be provided for each author, if available. Biographies should be placed at the end of the paper, after the references. Personal information such as hobbies or birthplace/birthdate should not be included. Author photographs are not published.

%\newpage

%\listoffigures

\end{spacing}
\end{document}